\documentclass[aip,two column]{revtex4}
\usepackage{stmaryrd}
\usepackage{mathrsfs}
\usepackage{amsfonts}
\usepackage[mathscr]{eucal}
\usepackage{amssymb}
\usepackage{amsmath}
\usepackage{dcolumn}
\usepackage{bm}
\usepackage{graphicx}

\begin{document}


\newcommand\ppcf{Plasma Phys. Control. Fusion }
\newcommand\pop{Phys. Plasmas }
\newcommand\nf{Nucl. Fusion }

\title{Finite-orbit-width effects on the geodesic acoustic mode in the toroidally rotating tokamak plasma}

\author{H. Ren}
\email{hjren@ustc.edu.cn}
\affiliation{CAS Key Laboratory of Geospace Environment and Department of Modern Physics, University of Science and Technology of China, Hefei 230026, P. R. China}


\date{submitted to \emph{Physics of Plasmas} on October 10, 2016}

\begin{abstract}
The Landau damping of geodesic acoustic mode (GAM) in a torodial rotating tokamak plasma is analytically investigated by taking into account the finite-orbit-width (FOW) resonance effect to the 3rd order. The analytical result is shown to agree well with the numerical solution. The dependence of the damping rate on the toroidal Mach number $M$ relies on $k_r \rho_i$. For sufficiently small $k_r \rho_i$, the damping rate monotonically decreases with $M$. For relatively large $k_r \rho_i$, the damping rate increases with $M$ until approaching the maximum and then decreases with $M$.
\end{abstract}



\maketitle
In the kinetic framework, keeping terms to the 1st order finite-Larmor-radius (FLR) effect, which represents the leading order polarization, and the 1st order finite-orbit-width (FOW) effect of passing particles $\delta_i \thicksim q \rho_i$, where $q$ is the safety factor of tokamaks and $\rho_i$ is the ion gyroradius, the dispersion relation of geodesic acoustic mode (GAM) \cite{winsor} is derived. The classical Landau damping rate of GAM is found to be $\propto e^{- q^2 \Omega^2}$ and independent of $\delta_i$, where $\Omega$ is the GAM frequency normalized by $R/v_{Ti}$ with major radius $R$ and ions thermal velocity $v_{Ti}$. Later, some theoretical analysis\cite{JPP2006, Nguyen2008,Zonca2008}, numerical evaluation \cite{Gaozhe2008}, and simulation \cite{Xu2008,Dorf2013} all indicate that the high-order FOW effect plays a key role in the collisionless damping of GAM, specifically in the large $q$ region. The resonant damping rate is sensitive to and significantly enhanced by $k_r \rho_i$, where $k_r$ is the radial wave number. When only the 2-nd resonance is taken into account, a recent NEMORB simulation \cite{NF2014} performed a good agreement with the analytical result to the 2-nd FOW effects for about $q < 3.5$\cite{JPP2006}. It was shown that for $q > 3.5$, the discrepancy between the theoretical result with 2nd harmonics and the TEMPEST simulation data becomes remarkable\cite{Xu2008}. That means higher-order resonance should be considered. The numerical evaluation performed by Gao {\it et al.} \cite{Gaozhe2008} shows that the damping rate with 3-rd resonance and the one with 4-th resonance have only slight discrepancy when $q$ is about greater than $7$. Meanwhile, Xu \emph{et al. } numerically found that the damping rate with 4-th order resonance is almost the same as the rate with 10-th resonance\cite{Xu2008}.

In a recent work\cite{Guo2015}, Guo and co-authors investigated the collisionless damping rate of GAM by taking into account the toroidal rotation. They numerically evaluated the influence of toroidal Mach number on the Landau damping rate by considering from 2nd to 5th FOW resonance effect, respectively. Similar to the case without toroidal rotation\cite{Gaozhe2008,Xu2008, HR2016}, they found that the damping rate was significantly enhanced by 3rd resonance and the damping rate with 3rd resonance and the one with 4th or 5th resonance are almost the same with each other. In this Letter, theoretical investigation on the collisionless damping of GAM in a toroidally rotating tokamak plasma is performed. We derive the analytical expression for the Landau damping rate by considering the FOW resonance effect to the 3rd order. Good agreement is found between our analytical result and the numerical evaluation. The $M$ scan of the damping rate is shown to significantly depend on $k_r \rho_i$.

Let us consider only the equilibrium toroidal rotation with $\vec{u}_0 = \omega_T(\psi) R^2 \nabla \zeta$. $\vec{w}$ is the particle velocity in the local reference frame moving with $\vec{u}_0$ relative to the lab frame. The perturbed distribution function is determined by the modified gyro kinetic (GK) equation applicable to low-frequency microinstabilities in a toroidally rotating tokamak, which shows\cite{GK,book,CGK1968}
\begin{align}
\delta F = q \delta \phi \bigg[ \frac{\partial F_0}{\partial E} + (1 - J_0^2(k_r \rho_j) ) \frac{1}{B} \frac{\partial F_0}{\partial \mu} \bigg] + J_0 \delta h,
\end{align}
in which $\delta h$  is governed by
\begin{align}
& \bigg[\frac{\partial }{\partial t} +  (w_\parallel \vec{b} + \vec{u}_0 + \vec{v}_D ) \cdot \nabla \bigg] \delta h\nonumber\\
= & - q J_0 \frac{\partial F_0}{\partial E} \bigg( \frac{\partial }{\partial t} + \vec{u}_0 \cdot \nabla \bigg) \delta \phi - J_0 \frac{q}{m_j \omega_c} \vec{b} \times \nabla \delta \phi \cdot \nabla F_0 \nonumber\\
+ & J_0 \frac{q}{\omega_c} \vec{b} \times \nabla \delta \phi \cdot [(w_\parallel \vec{b} + \vec{u}_0) \cdot \nabla \vec{u}_0 + \nabla \vec{u}_0 \cdot (w_\parallel \vec{b} + \vec{u}_0)] \frac{\partial F_0}{\partial E}.
\end{align}
Only perturbed electrostatic potential is taken into account, which is justified for electrostatic GAM in a low-$\beta$ plasma. Here, $\delta \phi$ is the perturbed electrostatic potential, $J_0$ is the zeroth-order Bessel function, $\rho_j = w_\perp/\omega_c$ is the Larmor radius, $\omega_c = q B/m_j$ is the gyro frequency, and $\vec{v}_D$ is the leading order drift velocity. In the local reference frame, the leading order electrostatic potential $\Phi_0$ is determined by $ e \Phi_0 = \frac{T_e}{1 + \tau} M^2$\cite{Hinton1985}. Here, $\tau = T_e/T_i$ is the temperature ratio. As a result, the drift velocity can be expressed as\cite{GK}
\begin{align}
&\vec{v}_D = [(w_\parallel^2 + w_\perp^2/2)/\omega_c] \vec{b} \times \nabla \ln B + \frac{\vec{b}}{\omega_c} \nonumber\\
& \times \bigg[\frac{q}{m_j} \nabla \Phi_0 + \vec{u}_0 \cdot \nabla \vec{u}_0 + w_\parallel (\vec{b} \cdot \nabla \vec{u}_0 + \vec{u}_0 \cdot \nabla \vec{b}) \bigg].
\end{align}
The velocity coordinate used here is $(E, \mu)$, in which the magnetic moment $\mu$ stands for $\frac{1}{2 B} m_j w_\perp^2$ and the energy $E$ is defined as $ E = \frac{1}{2} m_j w^2 - \frac{1}{2} m_j u_0^2 + q \Phi_0$. As a result, the bi-Maxwellian ions equilibrium distribution can be written as $ F_0^i (\psi, E) = N(\psi) (\pi v_{T i}^2)^{- \frac{3}{2}} e^{ - \frac{E}{T_i}}$.

We focus on the ions perturbed distribution function. Using the properties of $\vec{u}_0$, one can show that $\nabla \vec{u}_0 = \omega_T R (\nabla R \nabla \zeta - \nabla \zeta \nabla R) + R^2 \nabla \omega_T \nabla \zeta$ and $\vec{u}_0 \cdot \nabla \vec{b} = \vec{b} \cdot \nabla \vec{u}_0$. Recalling the expression of $\Phi_0$, we have
\begin{align}
\vec{v}_D \cdot \nabla \psi = & \,\frac{I B}{\omega_c} (w_\parallel^2 + w_\perp^2/2) \nabla_\parallel \bigg( \frac{1}{B} \bigg) + \frac{\omega_T^2 I}{2 \omega_c (1 + \tau)} \nabla_\parallel R^2\nonumber\\
&  + \frac{w_\parallel \omega_T B}{\omega_c} \nabla_\parallel R^2.
\end{align}
It is also found
\begin{align}
(w_\parallel \vec{b} & + \vec{u}_0) \cdot \nabla \vec{u}_0 + \nabla \vec{u}_0 \cdot (w_\parallel \vec{b} + \vec{u}_0) \nonumber\\
&  = \bigg( \frac{I w_\parallel}{B} + \omega_T R^2 \bigg) \nabla \omega_T.
\end{align}
The GK equation is then reduced to
\begin{align}
\partial_\theta \delta h - i n_d \sin \theta \delta h - i \frac{\omega}{\omega_t} \delta h = i e J_0 \frac{\omega}{\omega_t} \frac{\partial F_0}{\partial E} \delta \phi.
\end{align}
This equation is not the same as the one in Ref. \cite{Guo2015}. The term in the bracket on the right-hand side of Eq. (4) in Ref. \cite{Guo2015} should just be 1 but not $\left(1 + 2 k \rho_\parallel \frac{\omega_T}{\omega} \sin \theta \right)$. Here, $\omega_t = w_\parallel/(q R)$ is the modified transit frequency and $n_d = k_r \delta^i$, where $\delta^i$ is short for $\frac{1}{\omega_t \omega_c R} \bigg( w_\parallel^2 + \frac{1}{2} w_\perp^2 + \frac{M^2 v_{T \parallel}^2}{1 + \tau} + 2 w_\parallel v_{T \parallel} M \bigg)$. The disturbed distribution function can be easily solved as
\begin{align}
\label{ionp}
 \delta F  & =  [1 - J_0^2 (k_r \rho_i)] \bigg( \frac{\partial F_0}{\partial E} + \frac{1}{B} \frac{\partial F_0}{\partial \mu} \bigg) e \delta \phi + e J_0^2  \frac{\partial F_0}{\partial E}\nonumber\\
& \times \sum_{n, k} i^{n - k} J_{n + l - k} (n_d) J_l(n_d) \frac{(l - k) \delta \phi_n e^{i k \theta}}{l - k + \omega/\omega_t} .
\end{align}

Here and below, we are restricted to the case of $\tau \ll 1$. So the perturbed potential $\delta \phi = \sum_m \delta \phi_m(\psi) e^{i m \theta}$ is reduced to $\delta \phi_0$ and the quasi-neutrality condition is simplified to $\int d^3 v \delta F_0 = 0$. Taking into account the FOW resonance effect to the 3rd order, the dispersion relation is obtained as
\begin{align}
\label{tau0}
&\underbrace{\frac{1}{2}}_{\textrm{FLR}_1} - \underbrace{\frac{3 k^2}{16}}_{\textrm{FLR}_2} + \underbrace{\frac{q^2}{2 \zeta} I_c}_{\textrm{FOW}_1} - \underbrace{\frac{q^2 k^2}{4 \zeta} \mathcal{R}_3(\zeta)}_{\textrm{FLR}_1 - \textrm{FOW}_1} + \underbrace{\frac{3}{64 \zeta} q^2 k^4 \mathcal{R}_4(\zeta)}_{\textrm{FLR}_2 - \textrm{FOW}_1} \nonumber\\
 & - \underbrace{\frac{q^4 k^2}{8 \zeta^3} [\mathcal{R}_1(\zeta) - 2 \mathcal{R}_1(\zeta/2) ]}_{\textrm{FOW}_2} + \underbrace{\frac{q^4 k^4}{16 \zeta^3} [\mathcal{R}_2(\zeta) - 2 \mathcal{R}_2 (\zeta/2) ] }_{\textrm{FLR}_1 - \textrm{FOW}_2} \nonumber\\
 & + \underbrace{\frac{q^6 k^4}{384 \zeta^5} [5 \mathcal{R}_0(\zeta) + 81 \mathcal{R}_0 (\zeta/3) - 64 \mathcal{R}_0 (\zeta/2) ]}_{\textrm{FOW}_3} = 0.
\end{align}
in which $\zeta$ is short for $q \Omega$, $k$ stands for $k_r \rho_i$ for simplicity of notation, and
\begin{equation}
I_c(\zeta) = \, \mathcal{Z}_4(\zeta) +  (1 + 6 M^2) \mathcal{Z}_2 (\zeta) + \left(\frac{1}{2} + M^2 + M^4 \right) \mathcal{Z}(\zeta),\nonumber
\end{equation}
and
\begin{widetext}
\begin{align}
 \mathcal{R}_0(\zeta)  = & \,\mathcal{Z}_{12} (\zeta) + (3 + 66 M^2) \mathcal{Z}_{10} (\zeta) + \left( \frac{15}{2} + 135 M^2 + 495 M^4 \right) \mathcal{Z}_8(\zeta)\nonumber\\
 & + (15 + 210 M^2 + 630 M^4 + 924 M^6) \mathcal{Z}_6(\zeta) + \left( \frac{45}{2} + 225 M^2 + 525 M^4 + 630 M^6 + 495 M^8 \right) Z_4(\zeta)\nonumber\\
& + \left( \frac{45}{2} + 135 M^2 + 225 M^4 + 210 M^6 + 135 M^8 + 66 M^{10} \right) \mathcal{Z}_2 (\zeta)\nonumber\\
 & + \left( \frac{45}{4} + \frac{45}{2} M^2 + \frac{45}{2} M^4 + 15 M^6 + \frac{15}{2} M^8 + 3 M^{10} + M^{12} \right) \mathcal{Z}(\zeta),\nonumber\\
\mathcal{R}_1(\zeta) = & \, \mathcal{Z}_8(\zeta) + (2 + 28 M^2) \mathcal{Z}_6(\zeta)  + (3 + 30 M^2 + 70 M^4) \mathcal{Z}_4(\zeta) + (3 + 18 M^2 + 30 M^4 + 28 M^6) \mathcal{Z}_2(\zeta)\nonumber\\
&  + \left( \frac{3}{2} + 3 M^2 + 3 M^4 + 2 M^6 + M^8 \right) \mathcal{Z}(\zeta),\nonumber\\
\mathcal{R}_2(\zeta) = & \, \mathcal{Z}_8(\zeta) + (4 + 28 M^2) \mathcal{Z}_6(\zeta) + (9 + 60 M^2 + 70 M^4) \mathcal{Z}_4(\zeta) + (12 + 54 M^2 + 60 M^4 + 28 M^6) \mathcal{Z}_2 (\zeta)\nonumber\\
 &  + \left(15/2 + 12 M^2 + 9 M^4 + 4 M^6 + M^8 \right) \mathcal{Z}(\zeta),\nonumber\\
 \mathcal{R}_3 (\zeta) = & \, \mathcal{Z}_4(\zeta) + (2 + 6 M^2) \mathcal{Z}_2 (\zeta) + \left( 3/2 + 2 M^2 + M^4 \right) \mathcal{Z}(\zeta),\nonumber\\
\mathcal{R}_4(\zeta) = & \,  2 \mathcal{Z}_4 (\zeta) + (6 + 12 M^2) \mathcal{Z}_2 (\zeta) + (6 + 6 M^2 + 2 M^4) \mathcal{Z}(\zeta).\nonumber
\end{align}
\end{widetext}
Here, $\mathcal{Z}_n(\zeta) \equiv \frac{1}{\sqrt{\pi}} \int_L \frac{x^n e^{- x^2} }{x - \zeta} d x $ and $\mathcal{Z}_0$ is abbreviated as $\mathcal{Z}$, which is the so-called plasma dispersion function. The subscript of FLR and FOW indicates the order, for example, FLR$_1$ means the 1st order FLR effect. One should note that the FLR$_3$ term is neglected since it contributes only to the real part of GAM frequency by introducing modification on the order of $\mathcal{O}(k^4)$. On the other hand, due to the Landau damping, the FOW$_3$ term affects the damping rate dramatically. According to the dispersion relation \eqref{tau0}, numerical result can be easily evaluated to explore the effects of torodial rotation on the GAM Landau damping.

Asymptotically expanding $\mathcal{Z}(\zeta)$ by assuming $\zeta \gg 1$ in Eq. \eqref{tau0} leads to the following simplified dispersion relation:
\begin{align}
\label{dis}
  &1 - \frac{3}{8} k^2 - \frac{G_1}{\Omega^2} - \frac{G_2}{\Omega^4} + i \sqrt{\pi} e^{- q^2 \Omega^2} q^5 \Omega^3 \left(1 + \frac{1 + 6 M^2}{q^2 \Omega^2} \right) \nonumber\\
  & +  i k^2 \sqrt{\pi} e^{- \frac{q^2}{4} \Omega^2} q^7 \Omega^3 \frac{q^2 \Omega^2 + 8 + 112 M^2}{512} \nonumber\\
& + i k^4 q^{11} \Omega^5 \sqrt{\pi} \left( e^{- \frac{q^2}{9} \Omega^2} G_3 -  e^{- \frac{q^2}{4} \Omega^2} G_4 \right) = 0,
\end{align}
in which
\begin{align}
&G_1 = \frac{7}{4} + 4 M^2 + M^4 - k^2  \left( \frac{13}{8} + \frac{5}{2} M^2 + \frac{1}{2} M^4 \right),\nonumber\\
&G_2 = \frac{1}{q^2} \left( \frac{23}{8} + 5 M^2 + \frac{M^4}{2} \right) \nonumber\\
& +  \frac{k^2}{64} (747 + 4176 M^2 + 3384 M^4 + 768 M^6 + 48 M^8),\nonumber\\
& G_3 = \frac{ 2 q^2 \Omega^2 + 54 + 1188 M^2 +  \frac{1215  + 21870 M^2 + 80190 M^4}{q^2 \Omega^2}}{2519424},\nonumber\\
& G_4 = \frac{q^2 \Omega^2 + 12 + 264 M^2 + \frac{12}{q^2} +  \frac{120 + 2160 M^2 + 7920 M^4}{q^2 \Omega^2} }{12288}.\nonumber
\end{align}
The real part of the simplified dispersion relation \eqref{dis} yields the frequency of GAM to the order of $1/q^2$ and $k^2$ as
\begin{align}
\label{fre}
&\Omega^2_{\textrm{G}} = \left(\frac{7}{4} + 4 M^2 + M^4 \right) \bigg[ 1 + \frac{1}{q^2} \frac{46 + 80 M^2 + 8 M^4}{(7 + 16 M^2 + 4 M^4)^2} \nonumber\\
& + k^2 \frac{1277 + 7632 M^2 + 6104 M^4 + 1344 M^6 + 80 M^8}{8 (7 + 16 M^2 + 4 M^4)^2} \bigg].
\end{align}
Meanwhile, the Landau damping rate with FOW resonance effect to the 3rd order is given as
\begin{align}
\label{damping}
\gamma_d =&  - \frac{q^5 \Omega_G^6 \sqrt{\pi}}{2 (G_1 + 2 G_2/\Omega_G^2)} e^{- q^2 \Omega_G^2} \bigg[ 1 + \frac{1 + 6 M^2}{q^2 \Omega_G^2} \nonumber\\
  & +   k^2 q^2  e^{\frac{3}{4} q^2 \Omega^2} \frac{q^2 \Omega_G^2 + 8 + 112 M^2}{512} \nonumber\\
& +  k^4 q^6 \Omega_G^2  \left( e^{\frac{8}{9} q^2 \Omega_G^2} G_3 -  e^{\frac{3}{4} q^2 \Omega^2} G_4 \right) \bigg].
\end{align}
The previous results \cite{HR2015a} are recovered when zeroing $k$ in the two equations above.

\begin{figure}[h]
\setlength{\unitlength}{0.5cm}
\begin{center}
\begin{minipage}[t]{8 cm}
\includegraphics[width= 7 cm]{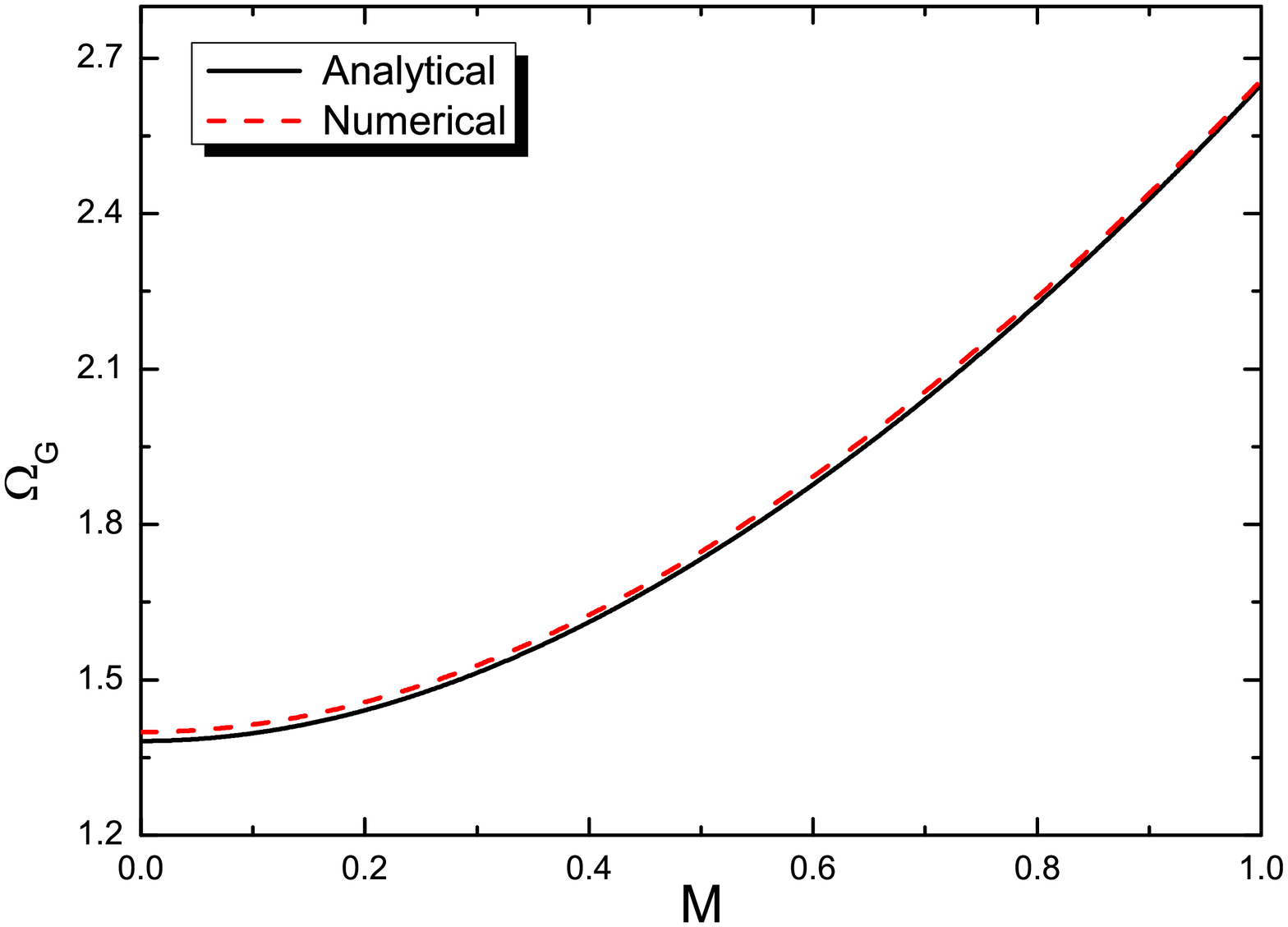}
\caption{The dependence of GAM frequency on the Mach number with $q = 4$ and $k = 0.1$. The solid curve is plotted according to the analytical frequency \eqref{fre} and the dashed one is plotted according to the exact numerical solution of Eq. \eqref{tau0}.}\label{fig1}
\end{minipage}
\end{center}
\end{figure}

\begin{figure}[h]
\setlength{\unitlength}{0.5cm}
\begin{center}
\begin{minipage}[t]{8 cm}
\includegraphics[width= 7 cm]{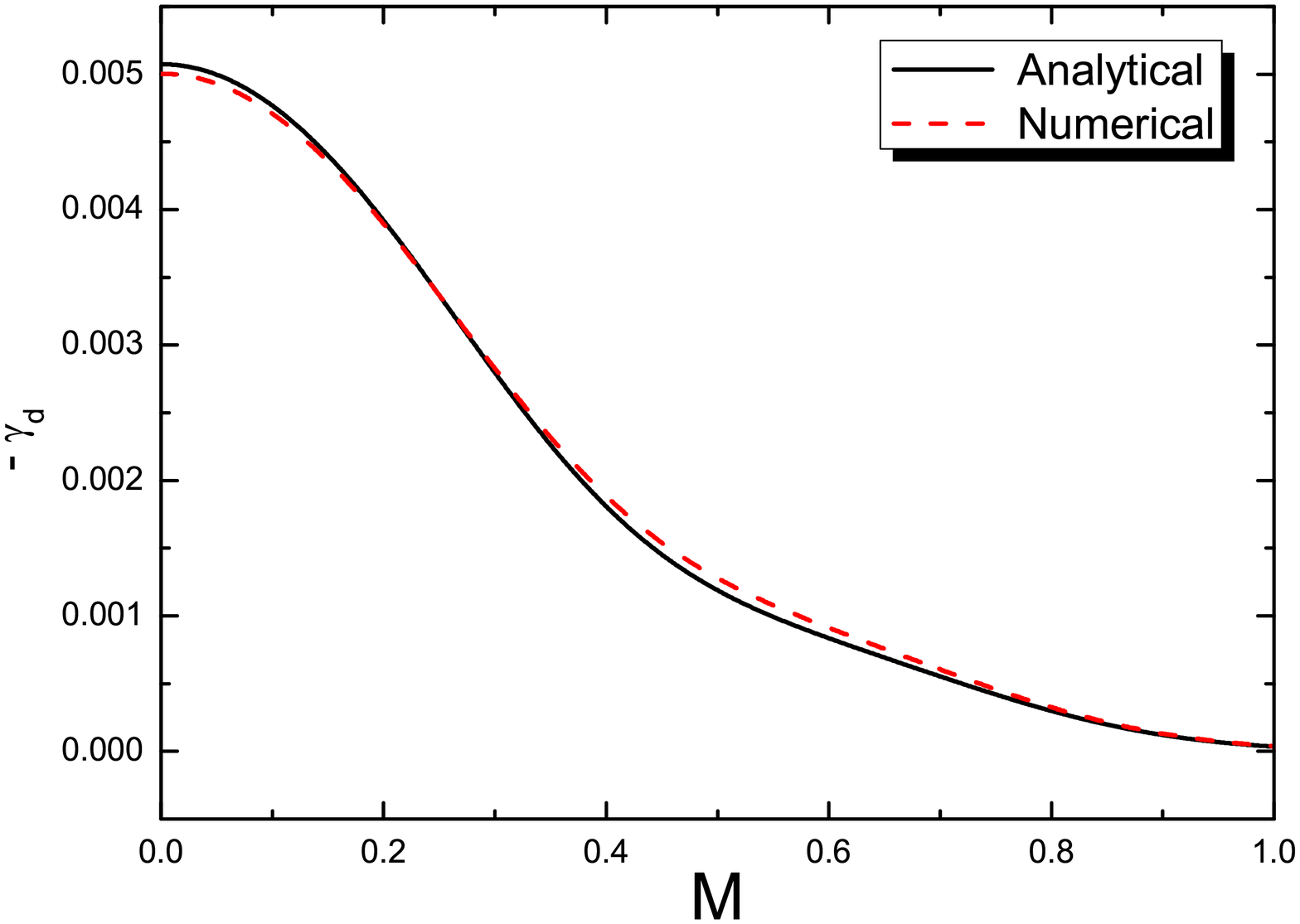}
\caption{The dependence of damping rate on the Mach number with $q = 4$ and $k = 0.05$. The solid curve is plotted according to the analytical damping rate \eqref{damping} and the dashed one is plotted according to the exact numerical solution of Eq. \eqref{tau0}. }\label{fig2}
\end{minipage}
\end{center}
\end{figure}

\begin{figure}[h]
\setlength{\unitlength}{0.5cm}
\begin{center}
\begin{minipage}[t]{8 cm}
\includegraphics[width= 7 cm]{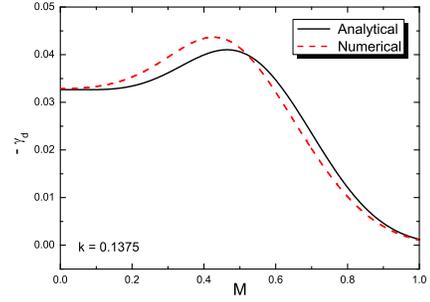}
\caption{The dependence of damping rate on the Mach number with $q = 4$ and $k = 0.1375$. The solid curve is plotted according to the analytical damping rate \eqref{damping} and the dashed one is plotted according to the exact numerical solution of Eq. \eqref{tau0}. }\label{fig3}
\end{minipage}
\end{center}
\end{figure}

Fig. \ref{fig1} illustrates the analytical frequency of GAM given in Eq. \eqref{fre} and the exact numerical solution of the dispersion relation \eqref{tau0} versus the Mach number $M$ for given $q = 4$ and $k = 0.1$. According to this figure, a perfect agreement is found between the analytical frequency and the numerical solution.
The dependence of collisionless damping rate of GAM on the Mach number is plotted in Fig. \ref{fig2}, where $q =  4$ and $k = 0.05$ is adopted by following Ref. \cite{Guo2015}. Quite different from Ref. \cite{Guo2015}, the damping rate with 3rd harmonics monotonically decreases with $M$ in this case. There is no fluctuations on the curves. As a comparison, we then let $k = 0.1375$ as done in the TEMPEST simulation \cite{Xu2008}, COGENT simulation\cite{Dorf2013} and theoretical calculation\cite{HR2016,Qiu2009}. It is found that in this condition, the damping rate increases with $M$ first. When $M$ is beyond a critical value, the damping rate goes to decrease as $M$ increases, as shown in Fig. \ref{fig3}. According to Figs. \ref{fig2} and \eqref{fig3}, one can see that the analytical result \eqref{damping} agrees well with the exact numerical solution.

\begin{figure}[h]
\setlength{\unitlength}{0.5cm}
\begin{center}
\begin{minipage}[t]{8 cm}
\includegraphics[width= 7 cm]{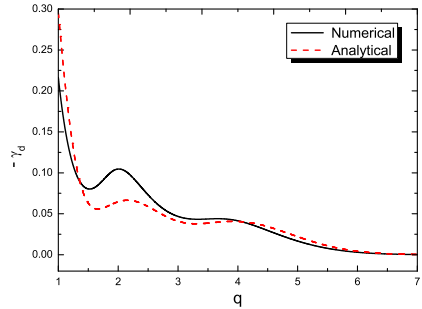}
\caption{The collisionless damping rate versus the safety factor $q$ with $M = 0.5$ and $k = 0.1375$. The dashed curve is plotted according to the analytical damping rate \eqref{damping} and the numerical curve (solid one) is plotted according to the exact numerical solution of Eq. \eqref{tau0}. The Mach number $M = 0.5$ is adopted. }\label{fig4}
\end{minipage}
\end{center}
\end{figure}

Furthermore, we plot the $q$ scan of the damping rate in Fig. \ref{fig4}. From this figure, we can see that there are two fluctuations on the curve. The damping rate decreases with $q$ at first. When $q$ is larger about $1.5$, the damping rate starts to increase with $q$ and then decreases with $q$ again when $q$ is about greater than $2$. The second fluctuation appears in the region $q \in (3.5, 4.5)$. Similar results have been reported in the previous simulation and theory study\cite{Gaozhe2008,Xu2008,Dorf2013,HR2016}. The first fluctuation is induced by the 2nd FOW resonance effect and the second fluctuation is mainly caused by the 3rd FOW resonance effect. Basically, the analytical result agrees with the numerical one for all systematic $q$ scan, especially for $q > 2.5$.

In summary, using modified gyro kinetic equation, we investigated the collisionless damping rate of GAM by considering the FOW resonance effect to the 3rd order harmonics in a toroidally rotating tokamak plasma. Asymptotically expanding the plasma dispersion function in the general dispersion relation \eqref{tau0} yields the simplified dispersion relation \eqref{dis}. According to the simplified one, we obtained the GAM frequency with modification to the order of $1/q^2$ and $k^2$ (see Eq. \eqref{fre}), and the analytical damping rate by keeping terms to the order of $k^4$ (see Eq. \eqref{damping}). Good agreement is found between the analytical results and the exact numerical solution of Eq. \eqref{tau0}. Fig. \ref{fig2} indicates that for $k = 0.05$, the damping rate monotonically decreases with $M$, while Fig. \ref{fig3} shows that for $k = 0.1375$, the damping rate increases with $M$ first. There exists a non-zero critical value of $M$ at which the growth rate becomes maximum. Beyond the critical value, the damping rate starts to decrease with $M$.

This work was supported by the China National Magnetic Confinement Fusion Energy Research Project under Grant Nos. 2015GB120005 and 2013GB112011, and the National
Natural Science Foundation of China under Grant No. 11675175.


\end{document}